Article

# Constraints on the Distribution of Gas and Young Stars in the Galactic Centre in the Context of Interpreting Gamma Ray Emission Features

**Steven N. Longmore [1],*** and J. M. Diederik Kruijssen [2]

[1] Astrophysics Research Institute, Liverpool John Moores University, 146 Brownlow Hill, Liverpool L3 5RF, UK
[2] Astronomisches Rechen-Institut, Zentrum für Astronomie der Universität Heidelberg, Mönchhofstraße 12-14, 69120 Heidelberg, Germany; kruijssen@uni-heidelberg.de
* Correspondence: s.n.longmore@ljmu.ac.uk

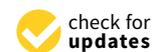

**Abstract:** Gamma ray observations have found evidence of an extremely energetic outflow emanating from the Galactic Centre, and an 'excess' of emission at GeV energies towards the Galactic Centre over that expected from current models. Determining whether the outflow is AGN- or star formation-driven, and whether the 'excess' is astrophysical in origin or requires new physics (e.g., self-annihilation of dark matter), requires the accurate modelling of the expected energy injection from astrophysical sources and the subsequent interaction with the surrounding environment. We briefly summarise current constraints on the distribution of gas and young stars in the inner few hundred parsecs of the Galaxy that can be included in future 2D and 3D modelling of the astrophysical gamma ray emission. The key points to highlight with respect to predominantly axisymmetric models currently in use are: (i) the distribution of dense gas, young stars and interstellar radiation field is highly asymmetric around the Galactic Centre; (ii) star formation is almost exclusively constrained to a Galactocentric radius of $\sim$100 pc; and (iii) the star formation rate in this region has been constant at $\lesssim$0.1 $M_\odot yr^{-1}$ to within a factor of 2 over the last $\sim$5 Myr.

**Keywords:** galactic centre; star formation; interstellar medium; Fermi bubble; galactic centre gamma ray excess

## 1. Introduction

Hosting our nearest supermassive black hole, sites of recent extreme star formation activity, and potentially large populations of compact accreting objects, the centre of our Galaxy is an ideal laboratory for studying high-energy physics [1]. Observations of the Galactic Centre from X-ray and gamma ray facilities (e.g., Fermi, HESS, Chandra, NuSTAR) are providing new insights into the energetic processes that shape the environment, probing the origin of cosmic rays, and even possibly the nature of dark matter through self-annihilation signals [2–8].

Given the implications for our understanding of mass flows and energy cycles in galactic nuclei and potentially even fundamental physics, major efforts are underway to determine the origin of these high-energy photons. Determining the astrophysical contribution to the high-energy photon distribution from different sources (e.g., star formation, supernovae, AGN episodes) relies on modelling the energy/momentum injection from relevant astrophysical sources and the subsequent interaction with the surrounding environment [9,10]. As well as accurate injection models, this requires knowing the relative locations of different injection sources and the variations in physical conditions (interstellar radiation field, gas density, gas temperature, magnetic field strength and morphology, etc.) across the Galactic Centre.

While the location of some injection sources are well constrained (e.g., the super-massive black hole, Sgr A*, and stars in the nuclear cluster), the distribution of other injection sources (e.g., young high-mass stars, supernovae) and the distribution of dense gas (the main reservoir of baryonic matter not trapped inside stars) are more uncertain. In this paper, we provide a brief overview of recent developments in our understanding of the relative distribution of dense gas and young stars. This is intended as an introduction to the current state of knowledge and to provide a feel for the inherent uncertainties rather than as an exhaustive review. Our primary goal is to focus on the implications of the distribution of dense gas and young stars for modelling the origin of high-energy photons in the Galactic Centre.

## 2. 2D Distribution of Dense (Molecular) Gas and Young Stars

As the solar system resides in the disc of the Galaxy, the first challenge in understanding the distribution of gas and young stars in the vicinity of the Galactic Centre is removing the contribution from foreground and background gas/stars. Figure 1 shows the distribution of dense gas (red) and young stars (blue) observed towards the inner few degrees in longitude of the Galaxy.

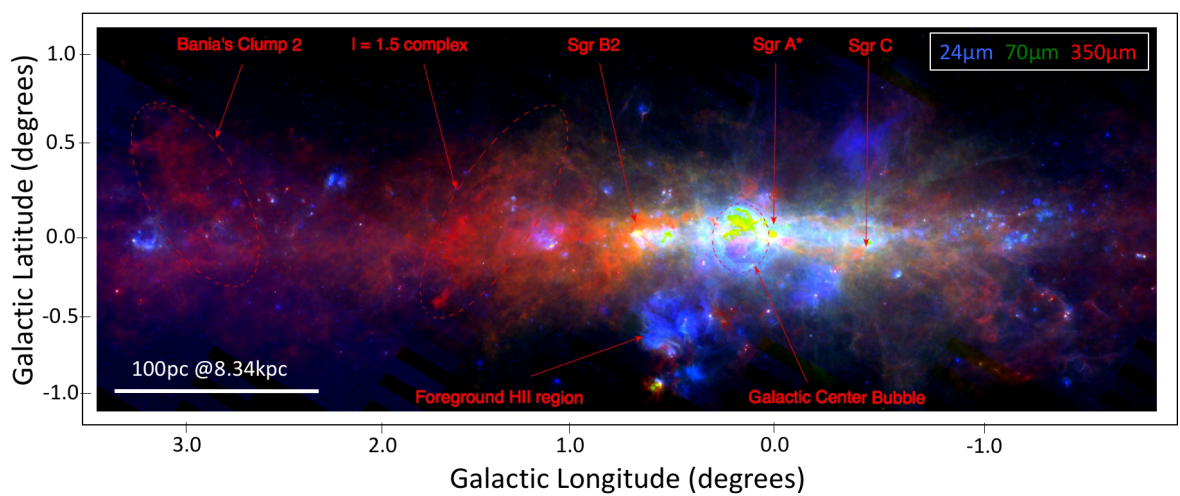

**Figure 1.** Three-colour image of the inner few degrees of the Milky Way around the Galactic Centre at sub-mm to mid-IR wavelengths (following [11,12]). Red and green show 350 µm and 70 µm emission (*Herschel* Hi-GAL [13]) and blue shows 24 µm emission (*Spitzer* MIPSGAL [14]). Sources of interest throughout the region are labelled. The scale-bar shows a projected length of ∼100 pc at a distance of 8.34 kpc [15]. As the largest concentration of dense gas in the Galaxy by far and the site of some of the most extreme star formations in the Galaxy, the inner few hundred pc of the Galactic Centre is very bright in (sub)-mm continuum emission from dust associated with the gas (red), and mid-IR emission from dust heated by recently formed high-mass stars (blue).

Thankfully, it is straightforward to tell that almost all of the emission from gas in the inner few degrees longitude lies within a few hundred pc of the Galactic Centre. This is because the properties of molecular gas in the Galactic Centre are very different from the gas in the disc [1,16,17]. Spectral line observations of molecular rotational transitions show that the gas velocity dispersion is many times larger at fixed spatial scale [18–23]. The molecular gas in the Galactic Centre is also on average a factor ∼100 times more dense and a factor 3–4 times warmer than gas in the disc [22,24,25]—conditions which excite more energetic rotational transitions that can be used to isolate gas at the Galactic Centre.

Similarly, it is possible to determine (e.g., from the measured extinction) that the vast majority of emission from young stars in the inner few degrees longitude must be coming from within a few hundred pc of the Galactic Centre. This region contains some of the most extreme recent star formations in the Galaxy. Sgr B2 is the most luminous star forming region in the Milky Way and often referred to as the closest example of a "mini-starburst" [26–30]. At a projected distance, $\sim$40 pc closer to the Galactic Centre than Sgr B2 (approximately at the location of the Galactic Centre Bubble annotation in Figure 1) lie the Arches and Quintuplet clusters, two of the most massive and dense young clusters in the Galaxy [31–33]. The nuclear cluster surrounding Sgr A* contains a population of very young and high mass stars (see [34] for a review). Finally, the very bright 24 µm (blue) point sources at negative Galactic longitudes in Figure 1 are a population of massive young stellar objects (YSOs) and evolved high mass stars [35–37].

One of the most striking features of the 2D projected distribution of the dense gas and young stars is their strong asymmetry: 2/3 of the dense gas is at positive longitudes and 2/3 of the young stars are at negative longitudes [1]. Numerical simulations suggest asymmetries in the dense gas may arise from unsteady inflow of gas in the barred potential (e.g., [38–41]).

## 3. 3D Distribution of Dense (Molecular) Gas and Young Stars

*3.1. 3D Distribution of Dense Gas*

The main challenge in transforming the 2D distribution of gas and young stars into a 3D distribution is ascertaining the distance to different emission regions. There are many different methods for attempting to determine distances, each of which have their strengths and weaknesses. These methods can be seperated into those that attempt to derive an absolute distance to an object, and those that try to infer the relative distances between objects.

Arguably the most robust absolute distance measurements come from cm-wavelength maser and continuum parallaxes. With multi-epoch VLBI observations, it is possible to fix the distance to bright objects (e.g., Sgr A* and masers in Sgr B2) with an uncertainty of a few percent [15]. With a model of the Galactic stellar population and the dust distribution along the line of sight to the Galactic Centre, it is also possible to derive an absolute distance to objects from near-IR extinction measurements. With an initial 3D framework tied down by these absolute distance measurements, it is possible to infer the relative position of different objects using other methods.

Combined with a model of the background stellar distribution (which dominates the gravitational potential), the change in the observed line-of-sight velocity of gas as a function of position provides an important handle on the location of the gas. From this, it has been inferred that the gas at positive longitudes (left) of Sgr B2 in Figure 1 is most likely at distances of a few hundred pc in the foreground, while the gas at negative longitudes (right) of Sgr C in Figure 1 is likely a few hundred pc in the background [42–44]. Unfortunately, current uncertainties in the distance mean we cannot localise the 3D position of any of this gas to better than a few hundred pc.

Fortunately, we have better constraints on the relative positions of gas and young stars in the inner 1 degree longitude (between Sgr B2 and Sgr C in Figure 1). Spectral line observations of molecular rotational transitions show that the majority of gas lies in clouds which are connected along a coherent gas stream [21,45–47]. Whether or not these clouds appear in absorption against the bright infrared background makes it possible to determine if they lie in front or behind the Galactic Centre. When combining this information with a model of the background stellar distribution (recall that it dominates the gravitational potential), one can derive the 3D distribution of the gas clouds from the change in the observed line-of-sight velocity of gas as a function of position [48,49]. Our current understanding is that most of the gas belongs to a stream of a few $10^7$ M$_\odot$ orbiting the centre at a radius of $\sim$100 pc [49]. Interior to the 100 pc gas stream, the $10^4$ M$_\odot$ 'circumnuclear disc' and the ionised gas 'mini-spiral' both lie within a few pc of Sgr A*.

A particularly interesting constraint on the distance of certain gas clouds from Sgr A* comes from spatial variations in the intensity of X-ray emission across these clouds as a function of time [50]. The interpretation of these variations is that they are X-ray light echoes, where the source of X-rays is assumed to be from Sgr A* when it had a flare in X-ray brightness, presumably from a prior accretion episode. With a model of the X-ray time variability of Sgr A*, one can infer the distance to clouds from the time lag since the outburst of their increasing X-ray brightness [51].

*3.2. 3D Distribution of Young Stars*

The nuclear star cluster, hosting a population of recently formed high-mass stars, is unambiguously situated in the inner pc of the galaxy surrounding Sgr A* [34]. The 3D location of the Arches and Quintuplet clusters is more difficult to ascertain. Measurement of their proper motion show they are moving with substantial velocity >100 km s$^{-1}$ to positive longitudes (left in Figure 1) along the Galactic plane, consistent with them orbiting the Galactic Centre at a radius of ∼100–200 pc [52–54]. Their orbital motions are consistent with them having formed in the 100 pc gas stream [49]. In the absence of distance and proper motion measurements, the 3D positions of the 24 μm point sources are relatively unconstrained. It has been postulated that these stars may be the remnants of an Arches/Quintuplet like cluster that has been destroyed by the tidal field (the tidal disruption timescale in this region of the Galaxy is 5–10 Myr: [55]). The fraction of young stars in bound clusters is difficult to ascertain due to incompleteness (observations are primarily sensitive to the youngest, most massive stars that only contain a small fraction of the young stellar mass budget), but appears close to 50%, consistent with theoretical predictions [56].

Almost all the current star formation activity and gas clouds just beginning to form are constrained to the 100 pc stream [30,57–70], as expected given the radial variation in environmental conditions within the inner few hundred pc of the Galaxy [71]. Observations that are sensitive to the star formation activity over different timescales show the rate of star formation has been constant to within a factor two of 0.1 M$_\odot$ yr$^{-1}$ for the last 5–10 Myr [72]. Although observations are unfortunately not sensitive to star formation rates on longer timescales, numerical simulations that resolve the large-scale (∼kpc) gas flows show the inflow is stochastic [38–41] and the star formation may be episodic with long periods (20–40 Myr) of quiescence (star formation rate ∼0.1 M$_\odot$ yr$^{-1}$) followed by intense (few Myr) periods where star formation activity is enhanced by an order of magnitude or more [73,74]. This is a consequence of the background stellar potential, environmental conditions, gas properties and the natural physical scales and time scales of the star formation process in the inner few hundred pc of the Galaxy [75–77].

## 4. Implications for Future Modelling of the Gamma Ray Emission

The distribution of gas and young stars provides several important constraints on the relative locations of energy injection sources, the surrounding gas reservoir, and the variation in physical conditions throughout the region. Specifically:

1. 2/3 of the dense gas is at positive Galactic longitudes (left of Sgr A* in Figure 1).
2. 2/3 of the young stars are at negative Galactic longitudes (right of Sgr A* in Figure 1).
3. All of the current star formations are contained within a projected radius of 100 pc from the Galactic Centre.
4. The star formation rate in this region has been constant at ∼0.1 M$_\odot$ yr$^{-1}$ to within a factor of 2 over the last ∼5 Myr. Numerical modelling suggests that the star formation rate may have been at last an order of magnitude larger in the past, but this is currently unconstrained by observations.

The implications of this for modelling the origin of high-energy photons are that the interstellar radiation field, mass reservoir, etc., are highly asymmetric and often anti-correlated on both large and small scales. These large-scale offsets suggest the distribution of core-collapse supernovae may also be asymmetric around the Galactic Centre. In summary, modelling (i) energy injection from young stars and core-collapse supernovae, and (ii) the variations in physical conditions (interstellar radiation field, gas density, temperature, magnetic field) as purely radial or axisymmetric terms should be treated with caution.

**Author Contributions:** Both authors contributed to writing and proof reading the article.

**Acknowledgments:** We thank the two anonymous reviewers for helpful feedback on the manuscript. SNL gratefully acknowledges funding support from STFC. JMDK gratefully acknowledges funding from the German Research Foundation (DFG) in the form of an Emmy Noether Research Group (grant number KR4801/1-1) and from the European Research Council (ERC) under the European Union's Horizon 2020 research and innovation programme via the ERC Starting Grant MUSTANG (grant agreement number 714907).

**Conflicts of Interest:** There are no conflicts of interest.